\documentclass[
reprint,
superscriptaddress,
showpacs,preprintnumbers,
nobibnotes,
 amsmath,amssymb,
aps,
]{revtex4-1}

\usepackage{graphicx}
\usepackage{dcolumn}
\usepackage{bm}
\usepackage[colorlinks,linkcolor=blue,citecolor=blue,urlcolor=blue,hyperindex,driverfallback=dvipdfm]{hyperref}
\usepackage[mathlines]{lineno}

\begin{document}


\title{Directional Generation of Graphene Plasmons by \\ Near Field Interference}

\author{Lei Wang}
\email{ggggcccf173@163.com}
\affiliation{College of Physics and Electronic Engineering, Xinyang Normal University, Xinyang 464000, China}%
\affiliation{The Key Laboratory of Weak-Light Nonlinear Photonics, Ministry of Education, School of Physics and TEDA Applied Physics Institute, Nankai University, Tianjin 300457, China}
\author{Wei Cai}
\email{weicai@nankai.edu.cn}
\author{Xinzheng Zhang}
\author{Jingjun Xu}
\email{jjxu@nankai.edu.cn}
\affiliation{The Key Laboratory of Weak-Light Nonlinear Photonics, Ministry of Education, School of Physics and TEDA Applied Physics Institute, Nankai University, Tianjin 300457, China}
\affiliation{Collaborative Innovation Center of Extreme Optics, Shanxi University, Taiyuan, Shanxi 030006, China}
\affiliation{Synergetic Innovation Center of Chemical Science and Engineering, Tianjin 300071, China}
\author{Yongsong Luo}
\affiliation{College of Physics and Electronic Engineering, Xinyang Normal University, Xinyang 464000, China}%

%
%

\date{\today}

\begin{abstract}
The highly unidirectional excitation of graphene plasmons (GPs) through near-field interference of orthogonally polarized dipoles is investigated. The preferred excitation direction of GPs by a single circularly polarized dipole can be simply understood with the angular momentum conservation law. Moreover, the propagation direction of GPs can be switched not only by changing the phase difference between dipoles, but also by placing the $z$-polarized dipole to its image position,  whereas the handedness of the background field remains the same. The unidirectional excitation of GPs can be extended directly into arc graphene surface as well. Furthermore, our proposal on directional generation of GPs can be realized in a semiconductor nanowire/graphene system, where a semiconductor nanowire can mimic the circularly polarized dipole when illuminated by two orthogonally polarized plane waves.
\end{abstract}

\maketitle


\section{\label{sec:level1}INTRODUCTION}

Graphene plasmons (GPs), the intrinsic collective oscillations of electrons coupled with electromagnetic waves in doped graphene, have attracted enormous interests for their unique properties, including such as inherently highly controllable, long-lived and extremely electromagnetic field confinement and enhancement in mid-infrared and terahertz spectral regimes\cite{LA2014, GPN2012, G2014}. Since theoretically proposed by Jablan \emph{et al} in 2009\cite{JBS2009}, GPs have been widely studied for electro-optical modulation\cite{LYU2011, yao2013broad}, quantum plasmonics\cite{tame2013}, light harvesting\cite{EBJ2011}, transformation optics\cite{VE2011} and infrared biosensors\cite{Rodrigo2015} at nanometer scale. Due to large wavevector mismatch between GPs and free light field, propagating GPs are usually excited by deep sub-wavelength point-like sources\cite{KCG2011}. In these cases, the sources such as  emitters, absorbers and scatterers serve as dipoles perpendicular or parallel to the propagation plane of GPs\cite{NGG2011}. However, the propagation direction of excited GPs is usually isotropic in the graphene plane as a result of the symmetry of structures and excitation configurations. And the unidirectional launching of GPs is still unsolved problem although it is important in ultracompact plasmonic devices at the chip scale. On the other hand, to achieve highly directional launching of surface waves in metal, attempts have been made to break the symmetry by introducing oblique incidence\cite{Lee2012, kim2009}, double slits\cite{li2011} and circularly polarized incident waves\cite{Rodriguez2013, xly2014}, etc. Among the numerous reported methods, the use of near field interference of a circularly polarized wave is cornerstone for active switching, along with very high extinction ratio between different directions. However, the current experimental effort to mimic a two-dimensional rotating dipole by oblique incidence is controversial due to the inevitable magnetic induction currents\cite{Lee2013c}. It is well-known that two orthogonally oriented dipoles can be induced by two orthogonally polarized incident plane waves. Aware of that a circularly polarized dipole can be efficiently mimicked by a nanowire illuminated by two orthogonal plane waves due to the extremely localization of GPs, it is natural for us to consider realizing unidirectional generation of GPs  by using a combined circularly polarized dipole in unstructured graphene. Furthermore, the combined dipoles can be separated in space compared to a single circularly polarized dipole, which provides us another degree of freedom to control the excitation of GPs.

In this study, two orthogonally oriented dipoles are employed to efficiently excite symmetric and anti-symmetric charge ordering modes in flat and arc graphene planes. As long as the constructive and destructive interferences of near-fields take place in different propagation directions, the unidirectional launching of GPs occurs. Due to the inherent phase difference between symmetric and anti-symmetric evanescent modes induced by $x$-polarized and $z$-polarized dipoles, the extra phase difference of $\pi$/2 , e.g., circularly polarized dipoles, should be introduced. Moreover, when the circularly polarized dipole is decomposed into two linear polarized ones and separated on different sides of the graphene, the behaviors of induced charge distribution and handedness of the background field are opposite. In further, the circularly polarized dipole can be efficiently mimicked by a semiconductor nanowire illuminated by two orthogonally polarized plane waves in experiments, and the problem about magnetic induction currents induced by oblique incidence can be solved in our considered system. We believe our findings should be found applications in compact plasmonic circuits in mid-infrared and terahertz regimes.\par
\section{\label{sec:level2}THEORETICAL BACKGROUND}
\begin{figure}[htb]
\includegraphics[width=8.5cm]{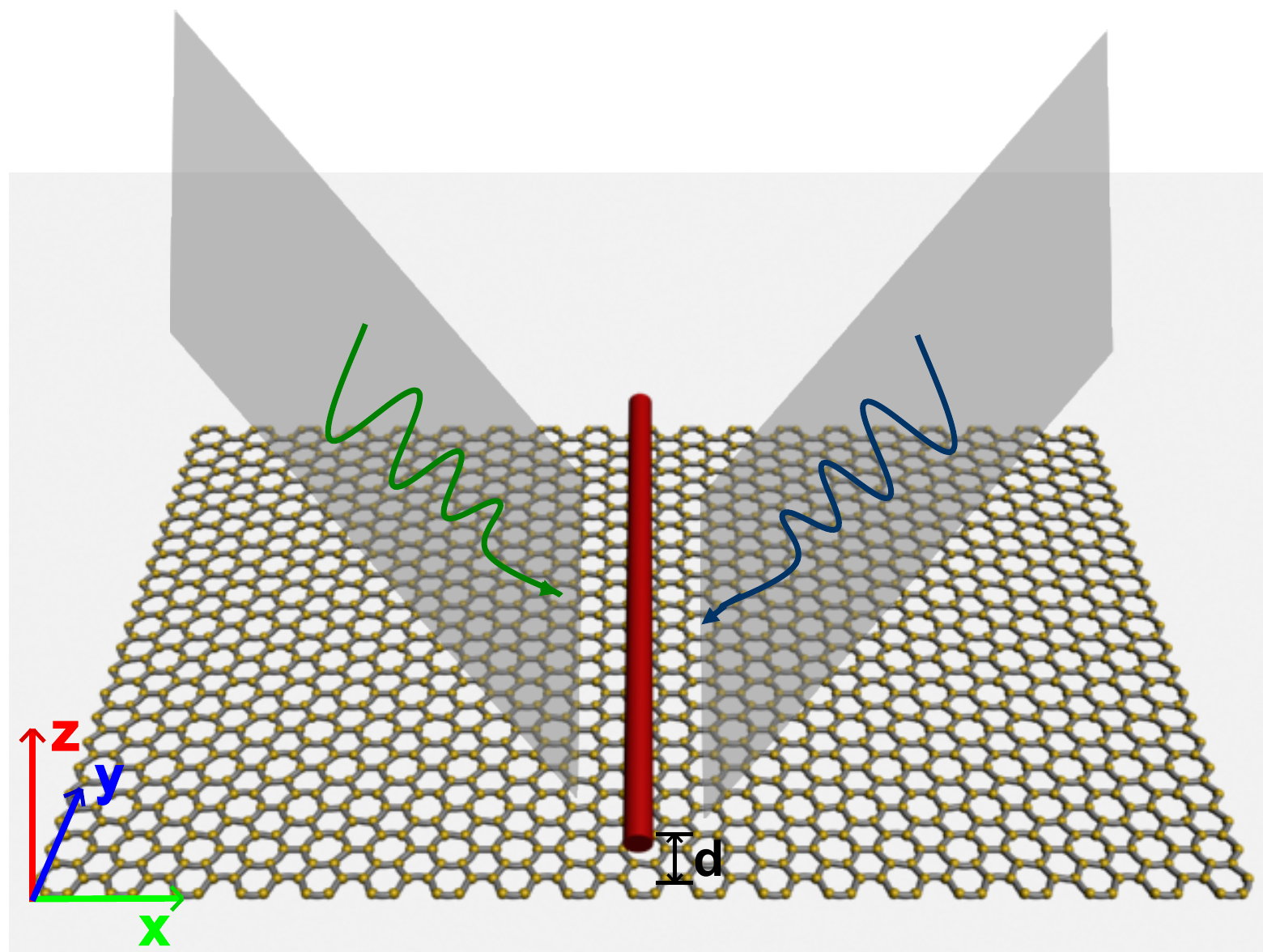}
\caption{\label{fig1} Schematic of directional excitation of graphene plasmons by a dipole source. The dipole source can be mimicked by a semiconductor nanowire. }
\end{figure}

The phenomena of unidirectional excitation of GPs can be understood by considering a dipole placed at a subwavelength distance close to a free standing graphene sheet. Fig. \ref{fig1}(a) illustrates the scheme employed in our design.  A two-dimensional (2D) dipole with momentum $\mathbf{p}_\text{2D}=[p_x, p_z]$ is placed above a graphene sheet. A Cartesian coordinate system is chosen with the graphene sheet laying in $z=0$ and the position of the dipole is (0, $z_{\text{dip}}$). Without loss of generality, the result can be extended to three-dimensional (3D) treatment directly\cite{Rodriguez2013}. The vector potential \textbf{A} induced by the dipole without graphene can be expressed as $\mathbf{A}(\mathbf{r})=-i\omega\mu_0G(\mathbf{r}',\mathbf{r})\mathbf{p}$, where $G(\mathbf{r},\mathbf{r}')=\frac{i}{4}H_{0}^{(1)}(k_0|\mathbf{r}-\mathbf{r}'|)$ is the 2D Green's function in free space\cite{novotny2012,tai1994}, and $k_0=\omega/c$ is wavenumber in vacuum. The angular spectrum decomposition of the vector potential can be written as
$\mathbf{A}(\mathbf{r})=\frac{\omega\mu_0\mathbf{p}}{4\pi} \int_{-\infty}^\infty{\frac{e^{ik_z|z-z_{\text{dip}}|}}{k_z}e^{ik_xx}dk_x}$, where $k_z=\sqrt{k_0^2-k_x^2}$ is the longitude wavenumber. Thus the magnetic field can be deduced as $H_y^0(x,z)=\frac{1}{\mu_0}(\nabla\times\mathbf{A})_y=\frac{i\omega }{4\pi}\int_{-\infty}^\infty{[p_z\frac{k_x}{k_z}\mp p_x]e^{ik_z|z-z_{\text{dip}}|}e^{ik_xx}dk_x}$. The angular spectrum of the magnetic field can be written as $H_y^0(k_x,z)=\frac{i\omega p_x}{4\pi}[\frac{p_z}{p_x}\frac{k_x}{k_z}\mp 1]e^{ik_z|z-z_{\text{dip}}|}$.

Next we turn to the case where the dipole is laid on top of graphene. Due to the extremely large wavenumber of GPs, i.e., $k_x=k_\text{spp} \gg k_0$, and $k_z=\sqrt{k_0^2-k_x^2}\approx ik_\text{spp}$ is satisfied,  so the induced magnetic field has a relation of $H_y^0(k_\text{spp},z)\propto \frac{p_z}{p_x}\mp i$. One can conclude that the complete interferences take place as long as $p_z$ and $p_x$ have equally modulus with phase difference of $\pm\pi$/2. The contribution from the graphene can be included by introducing reflected and transmitted fields, which can be calculated via simply multiplying the individual angular spectrums with corresponding Fresnel coefficients $r^p$ and $t^p$, respectively. When the dimensionless conductivity $\alpha=2\pi\sigma/c$ in units of the fine-structure constant $\alpha_0\simeq 1/137$ was adopted, the Fresnel coefficients can be written as\cite{KCG2011,NGG2011}:
\begin{eqnarray}\label{eq1}
  r^p(k_x) &=& 1-\frac{2k_z^{'}}{\epsilon_sk_z + k_z^{'} + 2\alpha\frac{k_zk_z^{'}}{k_0}} \\
  t^p(k_x) &=& \frac{2\epsilon_sk_z}{\epsilon_sk_z + k_z^{'} + 2\alpha\frac{k_zk_z^{'}}{k_0}}
\end{eqnarray}
where $\epsilon_s$ is dielectric constant of substrate, $k_z=\sqrt{k_0^2-k_x^2}$ and $k_z^{'}=\sqrt{\epsilon_sk_0^2-k_x^2}$ are magnitudes of the longitudinal wavenumbers. For the reflected and transmitted fields, one can obtain the angular spectra $H_y^{\text{ref}}(k_x,z)=r^p(k_x)\frac{i\omega p_x}{4\pi}[\frac{p_z}{p_x}\frac{k_x}{k_z}+1]e^{ik_z(z+z_{\text{dip}})}$ and $H_y^{\text{tr}}(k_x,z)=t^p(k_x)\frac{i\omega p_x}{4\pi}[\frac{p_z}{p_x}\frac{k_x}{k_z}+1]e^{-ik_z(z-z_{\text{dip}})}$, respectively. The angular spectra of total magnetic fields in the spaces on top of and bottom of graphene are calculated as $H_y(k_x,z)=H_y^0(k_x,z)+H_y^{\text{ref}}(k_x,z)$ and $H_y(k_x,z)=H_y^{\text{tr}}(k_x,z)$, respectively. From the customary boundary condition and charge conservation law, i.e.,  $n\times(\mathbf{H}_2-\mathbf{H}_1)=\mathbf{K}=\sigma \mathbf{E}_{\|}$ and $\nabla_s\cdot \mathbf{K}=i\omega\rho_s$, the induced charge density $\rho_s^{ind}$ in the graphene layer can be obtained from the difference of magnetic fields at each side of the graphene
\begin{eqnarray}\label{rho}
 \nonumber   \rho_s^{ind}(x)&=&\frac{\delta(z)}{i\omega}\frac{\partial~\sigma E_x(x,0)}{\partial x}\\
    &=&\frac{\delta(z)}{i\omega}\frac{\partial}{\partial x}(H_y(x,0^-)-H_y(x,0^+)).
\end{eqnarray}
From now on, the prefactor $\delta(z)$ will be omitted for convenience, thus the angular spectrum of $\rho_s^{ind}$ can be written as
\begin{equation}\label{rho}
    \rho_s^{ind}(k_x)={[t^p-(1+r^p)]}\frac{ik_x}{4\pi}[p_z\frac{k_x}{k_z}+p_x]e^{ik_zz_{\text{dip}}}.
\end{equation}

Noting that $r^p$ and $t^p$ depend on the modulus of $k_x$ only, naturally, one can divide the contributions of a circularly polarized dipole into two parts, i.e., $\rho_s^{ind}(k_x)=\rho^{p_z}(k_x)+\rho^{p_x}(k_x)$. The former induced by $z$-polarized dipole (abbreviated as $p_z$ for convenience) satisfies $\rho^{p_z}(k_x)\propto p_z\frac{k_x^2}{k_z}$, while the latter induced by $x$-polarized dipole satisfies $\rho^{p_x}(k_x)\propto p_xk_x$. The fundamental mechanism for directional generation is the charge density induced by $p_z$ has an even parity both in angular spectrum and real space\cite{sup}, whereas the opposite hold true for a $p_x$ dipole. The superposition of $\rho^{p_z}$ and $\rho^{p_x}$ (not the $H_y$ with opposite parities) leads to the constructive and destructive interferences in different directions.

When the $p_z$ or $p_x$ dipole is moved to its image position (0, -$z_\text{dip}$), noting that $\tilde{H}_y(k_x,0^-)-\tilde{H}_y(k_x,0^+)=H_0(k_x,0)(1+r^p-t^p)=H_y(k_x,0^+)-H_y(k_x,0^-)$ and the minus sign should be adopted before $p_x$ in the expressions of $H_y$ thanks to $z-(-z_\text{dip})>0$, therefore the charge density can be written as
\begin{eqnarray}\label{rhotilde}
 \nonumber \tilde{\rho}^{p_z}(k_x)&=&{(1+r^p-t^p)}\frac{ip_z}{4\pi}\frac{k_x^2}{k_z}e^{ik_zz_{\text{dip}}}=-\rho^{p_z}(k_x),\\
  \tilde{\rho}^{p_x}(k_x)&=&{(1+r^p-t^p)}\frac{i(-p_x)}{4\pi}{k_x}e^{ik_zz_{\text{dip}}}=\rho^{p_x}(k_x),
 \end{eqnarray}
where the tildes means the quantity induced by dipoles located at its image position. Therefore, one can obtain the relation of $\tilde{\rho}^{p_z}(x)=-\rho^{p_z}(x)$ and $\tilde{\rho}^{p_x}(x)=\rho^{p_x}(x)$. This result means that moving the $p_z$ dipole to its image position will switch the preferred propagation direction of GPs, while moving the $p_x$ dipole will not change the preferred direction. This behavior is quite counterintuitive. Because the magnetic field induced by the dipole satisfied $\tilde{H}_y^{p_z}=H_y^{p_z}$ and $\tilde{H}_y^{p_x}=-H_y^{p_x}$, which means that moving a $p_x$ dipole to its image position will change the incident dipole fields, while they keep unchanged when moving $p_z$ dipoles. Combination of these two facts leads to an amazing result that the incident and induced fields have different preferred propagation directions. Remarkably, the finally preferred direction of GPs is determined by the induced charge pattern rather than the incident field.

There are a lot of parameters to quantitative describe the asymmetrical excitation. Among them, the angular spectrum ratios satisfy $R_k[F]\equiv|F(k_x)/F(-k_x)|=|\frac{\frac{p_z}{p_x}\frac{k_x}{k_z}+1}{-\frac{p_z}{p_x}\frac{k_x}{k_z}+1}|$, ($F\in\{H_y, E_z, E_x, \rho\}$) which depend on $\frac{p_z}{p_x}\frac{k_x}{k_z}$ only. Moreover, the spatial dependent near field ratios defined as $R_x[F]\equiv|F(x)/F(-x)|$, ($F\in\{H_y, \rho, P_x\}$) are also very important for GPs which can be obtained via full-field simulations and verified by near field experiments directly. In engineering, another important parameter to quantify the asymmetrical transmission is extinction ratio, which is defined as the logarithm of energy flux ratio in opposite directions $\eta=10\log(P_r/P_l)$. The right and left energy flux along the graphene can be obtained by integrating the relative Poynting vector along $z$ direction far from the dipole source
\begin{equation}
   P_x(x\rightarrow\infty)=\int_{-\infty}^{\infty}\frac{1}{2}\text{Re}\{E_yH^*_z\}dz.
\end{equation}

\section{\label{sec:level3}NUMERICAL SIMULATIONS}
We present several scenarios in which the proper choices of dipoles close to graphene sheet provide possibilities for directional excitation of GPs. First of all, we consider the basic model described in Sec \ref{sec:level2}, and compare the simulated results to the analytical results calculated from angular spectra. In the simulation, the frequency of electromagnetic field emitted by the dipole is 30~THz (corresponding to $\lambda\approx 10~\mu m$). The dipole is situated at a distance of $d = 0.01\lambda$ on top of the graphene plane and has a momentum of [1, $\frac{p_z}{p_x}$]$p_x$, where the unit length momentum $\dot{p}_x=-i\omega p_x=1 \text{A}\cdot \text{m}$ is adopted for convenience, and the ratio $\frac{p_z}{p_x}$ is discussed later. Chemical potential of the doped graphene is set as $\mu$=0.4 eV, and ambient temperature is set as T=300 K, the in-plane conductivity of the graphene is computed within the local-random phase approximation (RPA)\cite{Gusynin2006,Falkovsky2007} with an intrinsic relaxation time $\tau=120$ fs (indicating the mobility of $\mu=3000$ $\rm{cm}^2$/Vs), which is a typical parameter derived from experiments\cite{cba2012,Rodrigo2015}. Commercial software COMSOL Multiphysics based on FEM method is adopted to solve the Maxwell equations. From the dispersion relations of GPs, the wavenumber of GPs for this free-standing graphene is $k_\text{spp}=\sqrt{1-\alpha^2}~k_0\approx21.78~k_0$, indicating the plasmon wavelength $\lambda_\text{spp}$ is 459~nm, and the extinction parameter is $a^*\equiv\frac{k_z}{k_x}|^{ }_{k_x=\text{Re}\{k_\text{spp}\}} =0.9989i$. Therefor $\frac{p_z}{p_x}=\pm 0.9989i$ will lead to completely destructive interference of the excited GPs in a certain direction.

\begin{figure}[htb]
\includegraphics[width=8.5cm]{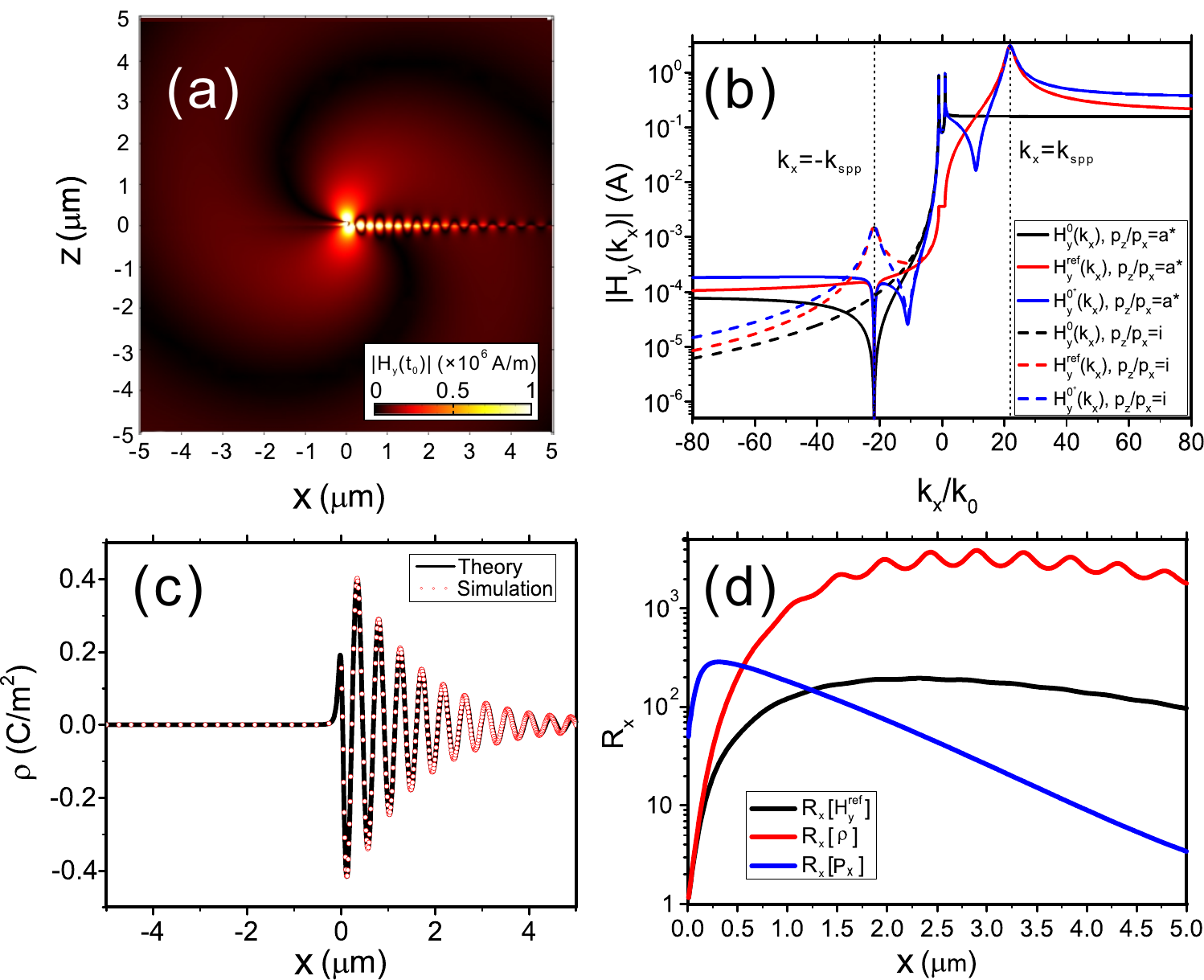}
\caption{\label{fig2}Directional excitation of GPs by a circularly polarized dipole. (a) Magnetic field distributions for GPs excited by a 2D circularly polarized dipole
$\mathbf{p}_{2D}=[1, a^*]p_0$, where $-i\omega p_0=1 \text{A}\cdot \text{m}$. (b) Angular momentum spectra of initial~($H_y^0$), reflective~($H_y^\text{ref}$) and total~($H_y$) magnetic field magnitude of the polarized dipole (solid lines) and an ideal circularly polarized (dashed lines). The dotted lines indicate the wavenumbers of GPs. (c) The simulated (solid line) and analytically calculated (marked by red circle) spatial dependent charge density in graphene. (d) Spatial dependent near field ratio $R_x[H_y^\text{ref}]=|H_y^\text{ref}(x)/H_y^\text{ref}(-x)|$(colored in black), $R_x[\rho]=|\rho(x)/\rho(-x)|$(colored in red) and $R_x[P_x]=|P_x(x)/P_x(-x)|$(colored in blue).}
\end{figure}
The simulated distribution of $H_y$ field $|\text{Re}\{H_y\}|$ is depicted in Fig. \ref{fig2}(a). Clearly, the plasmon mode is unidirectionally excited by the circularly polarized dipole, which has a much larger amplitude along $+x$ than $-x$. Besides, one can find that the background field is anticlockwise rotational due to the individual rotational dipole source. Thus  the angular momentum density of the background field is $\mathbf{L}=\epsilon_0\mathbf{r}\times (\mathbf{E}\times\mathbf{B})=\frac{1}{c^2}\mathbf{r}\times\mathbf{S}$, which is along $-y$ direction, where $\mathbf{S}$ is the Poynting vector\cite{jackson1962, EN1994}. Moreover, the angular momentum direction of preferred exited GPs is consistent with the angular momentum direction of background fields due to the conservation of angular momentum. When the phase difference between $p_x$ and $p_z$ changes from $\pi/2$ to $-\pi/2$ or placing the rotational polarized dipole on bottom of the graphene plane, the directions of angular momentum as well as the propagating GPs inverse. Therefor one can determine the preferred excitation directions simply by the direction of angular momentum. On the other hand, the directional excitation of GPs in real space can be understood by the asymmetry of the angular spectra in different directions. The angular spectra of initial, reflected and total magnetic fields are shown in Fig. \ref{fig2}(b). One can find the angular spectra have constructive and destructive interferences at $+k_x$ and $-k_x$, respectively. To understand this effect, the Fresnel coefficient $r^p$ is considered within plasmon pole approximation ( $k_x\simeq\pm k_\text{spp}$). The coefficient $r^p\propto\frac{1}{k^2_x-k^2_\text{spp}}$ has peaks when $k_x=\pm \text{Re}\{k_\text{spp}\}$. Specifically, when $\frac{p_z}{p_x}$ equals to $a^*$, $H_y^0(k_\text{spp},0^+)\propto \frac{p_z}{p_x}\frac{k_x}{k_z}|_{k_x=\text{Re}\{k_\text{spp}\}}+ 1=2$, this is a peak due to constructive interference and the intensity is twice larger than the magnitude of magnetic field induced by $p_x$ or $p_z$ individually. In the same manner, the valley exists due to constructive interference occurs at $|H_y^0(-k_\text{spp},0^+)|=0$. As a result, the angular spectra of initial, reflected and total fields have peaks at $+k_\text{spp}$ as well as valleys at $-k_\text{spp}$ for $\frac{p_z}{p_x}=a^*$. As to an ideal rotational polarized dipole, e.g. $\frac{p_z}{p_x}$ equals to $1i$, and $H_y^0(k_\text{spp},0^+)\propto \frac{k_x}{k_z}|_{k_x=\text{Re}\{k_\text{spp}\}}+ 1$ is about 2 owing to $a^*\approx 1i$. However,  there is a remarkable difference near $k_x=-k_\text{spp}$, where it is a peak instead of valley for the ideal circularly polarized dipole.  This comes from that  $|r^p|$ is maxima at $-k_\text{spp}$ and $|H_y^0(-k_\text{spp},0^+)|\propto|-i/a^*+1|$ is a slowly varying quantity.  In further, the spatial distribution of charge density is a vital physical quantity to describe the collective oscillations, such as plasmons. In Fig. \ref{fig2}(c), the simulated charge density distribution in the graphene plane is compared to the analytical result from  Eq. (\ref{rho}). They are in perfect agreement and have apparently constructive and destructive interferences in $x>0$ and $x<0$, respectively. The spatial dependent near field ratios of exited GPs are shown in Fig. \ref{fig2}(d). We can see that the near field ratios of reflected magnetic, charge density and energy flux are over 100 for $x<3\lambda_\text{spp}$.

\begin{figure}[htb]
\includegraphics[width=8.5cm]{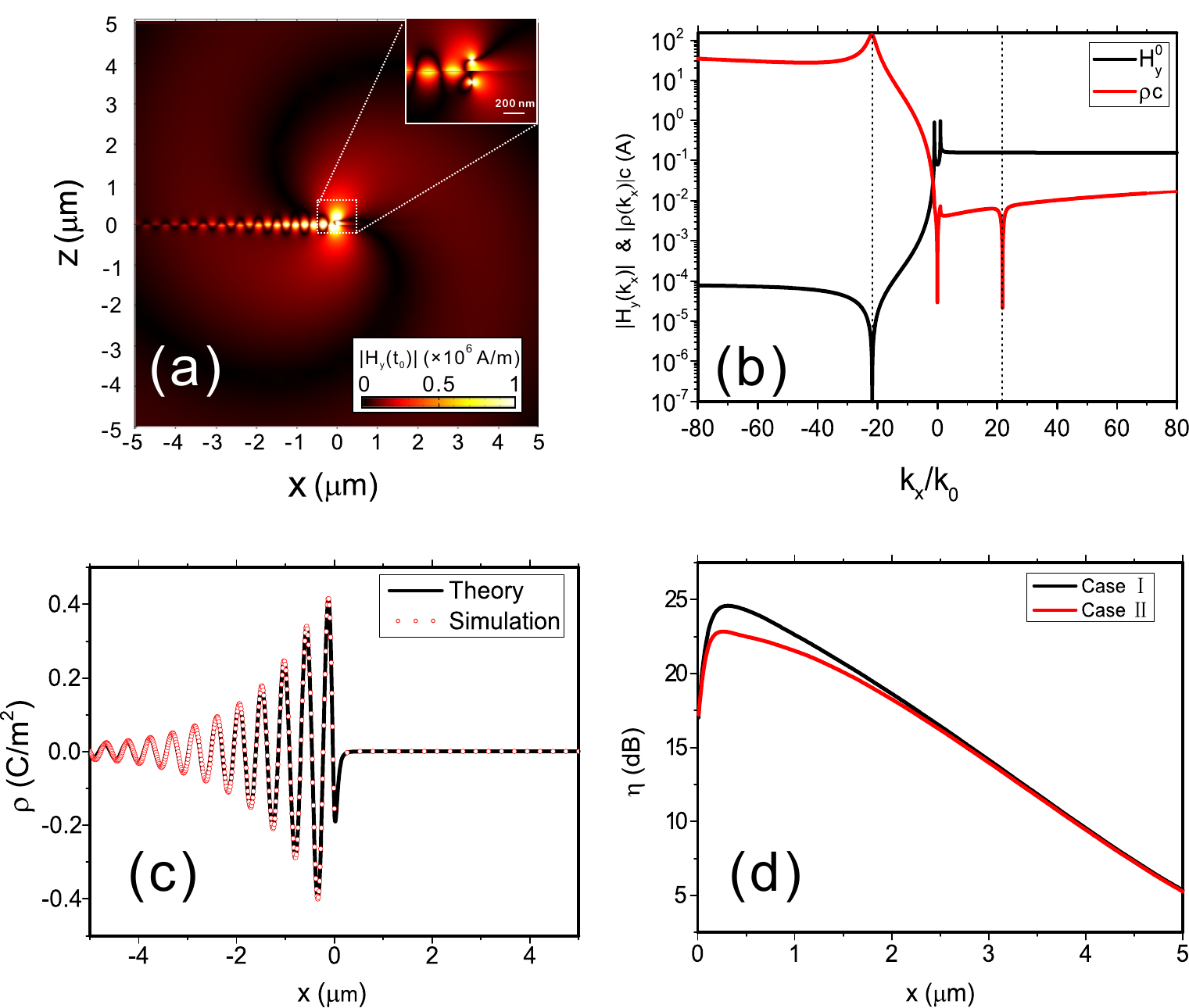}
\caption{\label{fig3} Directional excitation of GPs by two separated orthogonal polarized dipoles. (a) Magnetic field distributions for GPs excited by two orthogonal polarized dipoles $p_x=[1, 0]p_0$ and $p_z=[0, a^*]p_0$ located at (0, 0.01$\lambda$) and (0, -0.01$\lambda$), respectively. The insert figure shows the  enlarged excitation region. (b) Angular momentum spectra of initial magnetic field magnitude~($H_y^0$) and charge density~($\rho$) in the graphene plane, the initial and induced quantities have opposite preferred direction for excitation of GPs. (c) The simulated (solid line) and analytically calculated (marked by red circle) spatial dependent charge density in graphene. (d) The spatial dependent extinction ratio for a single circularly polarized dipole (case \uppercase\expandafter{\romannumeral1}, colored in black), and two separated orthogonal polarized dipoles (case \uppercase\expandafter{\romannumeral2}, colored in red).}
\end{figure}
A circularly polarized dipole can be decomposed into two orthogonal polarized dipoles, thus it is interesting to see what will happen if these two dipoles are placed on the different sides of graphene. Specifically, two orthogonal polarized dipoles with momenta $p_x=[1,0]p_0$ and $p_z=[0, a^*]p_0$ are considered, where they are located at (0, 0.01$\lambda$) and (0, -0.01$\lambda$), respectively. The simulated results are shown in Fig. (\ref{fig3}). One can find that the rotational direction of background field in Fig. \ref{fig3}(a) is the same as that in Fig. \ref{fig2}(a), however, the propagation direction of excited GPs inverses. This result means that one cannot distinguish these two cases from the far fields excepted for the preferred propagation direction of excited GPs. Considering that the $H_y$ is discontinuous across the graphene plane, the $H_y^\text{ref}$  changes from upside to downside of graphene when the dipole is placed to its image position. Meanwhile, $H_y^0$ and the induced charge $\rho$ remain unique. Similar to the results in Fig. \ref{fig2}(b), the angular spectra of $H_y^0$ and $\rho$ are also shown in Fig. \ref{fig3}(b) to demonstrate the mechanism of directional excitation in this scenario. Remarkably, the preferred directions of background field and induced charge are opposite, which are along $+x$ and $-x$, respectively. This result is rather counterintuitive and means that the angular momentum is 'non-conservation' at first sight. Actually, one can find this puzzling result comes from the magnetic field discontinuity at upper and lower sides of graphene ($r\neq t-1$), and the angular momentum is conservation.  In our proposed scheme, the angular momentum is zero at the original point with both positive and negative signs in $z=0$ plane simultaneously. Thus the use of the conservation of angular momentum cannot determine the preferred direction of GPs directly. To demonstrate the physical factor to determine the preferred directions, we turn to see the dependence of directional generation on phase difference $\Delta\phi=\phi^{p_z}-\phi^{p_x}=\arg\{\frac{p_z}{p_x}\}$. Similar to the superposition of polarizations, these two opposite sense of rotations lead to a classification of vibration ellipses according to their handedness, which is decided by the phase difference of two vibration vectors. If the phase difference satisfies $\Delta\phi=m\pi, m=0,1,2\cdots$, the superpositions are linear polarized dipoles, and their angular momenta is zero due to $\mathbf{r}//\mathbf{S}$, thus the excited GPs should be isotropic without any other asymmetry to fulfill the conservation of angular momentum. When the phase difference satisfies $\Delta\phi=\pi/2\pm2m\pi$, the near fields rotates in the anticlockwise sense, it is said to be left-handed. If extra $\pi$ phase is introduce to the $\Delta\phi$, the handedness and preferred direction of excitation will change. In the proposed system, there are three important factors to determine the handedness and preferred direction. The first one is the initial phase difference $\Delta\phi$ which is from the dipoles themselves, e.g., if the  initial $\arg\{\frac{p_z}{p_x}\}$ changes from $\pi/2$ to $\pi/2\pm\pi$, the handedness, i.e., rotational direction of background field and preferred direction will change. The second factor is the dipole position relative to the graphene. From the relation $H_y=H_y^{p_z}+H_y^{p_x}=i/\mu[k_x A_z+\textbf{sgn}(z_\text{dip}-z)k_z A_x]$, one can known that moving $p_x$ dipole to its image position will introduced a minus sign due to the sign function, which is equivalent to introduce extra $\pi$ phase difference when talking about the handedness and the initial electromagnetic field, while there is no extra phase difference when moving the $p_z$ dipoles to their image position. The total extra phase difference from aforementioned two factors will determine the handedness and the preferred direction of initial field. However, they are insufficient to determine the preferred direction of exited GPs. Noting that the scattering field of upper and lower sides of graphene satisfied $-r=(t-1)$ for free standing graphene, which introduced a minus sign compared to the continuous boundary condition $r=t-1$. This is the last vital factor to determine the preferred direction of induced field. Actually, the aforementioned counterintuitive result is originated from the minus sign, which can not be treated as extra phase difference as before because it only acts on induced field and do not affect the handedness and the distribution of initial magnetic field. In a word, there are three factors for $p_x$ to affect the preferred direction of excited GPs, while only two factors for $p_z$ to affect the preferred direction of excited GPs in our considered system. When the $p_x$ and $p_z$ locate in the same side of graphene, $H_y^\text{ref}(z=0)$ always denotes the magnetic field in the dipole side, thus the angular momentums of $H_y^0$ and $\rho$ have the same sign owing to the conservation of angular momentum. When they are in the different side, $H_y^\text{ref}$ in $z=0$ induced by the two dipoles denotes different sides of graphene, and the preferred direction of GPs should be decided by $\rho$ rather than $H_y^0$. Due to the extra minus sign, the preferrer direction of $\rho$ and $H_y^0$ is always opposite in this condition. The spatial dependent induced charge density is plotted in Fig. \ref{fig3}(c). One can find that the charge oscillates only in the -$x$ direction which is in good agreement with the analytical result. The comparison of spatial dependent extinction ratio of a single circularly polarized dipole (named after case \uppercase\expandafter{\romannumeral1}) and two orthogonal polarized dipoles placed at both sides of the graphene (named after case \uppercase\expandafter{\romannumeral2}) are shown in Fig. \ref{fig3}(d). One can see the extinction ratio is over 20 for $x< 4\lambda_\text{spp}$, and the ratio in case \uppercase\expandafter{\romannumeral2} is less than that in case \uppercase\expandafter{\romannumeral1}, this originates from the opposite preferred directions of the initial and induced fields in case \uppercase\expandafter{\romannumeral2}. The difference on extinction ratio between these two cases can be ignored when $x\gtrsim5\lambda_\text{spp}$.

\begin{figure}[htb]
\includegraphics[width=8.5cm]{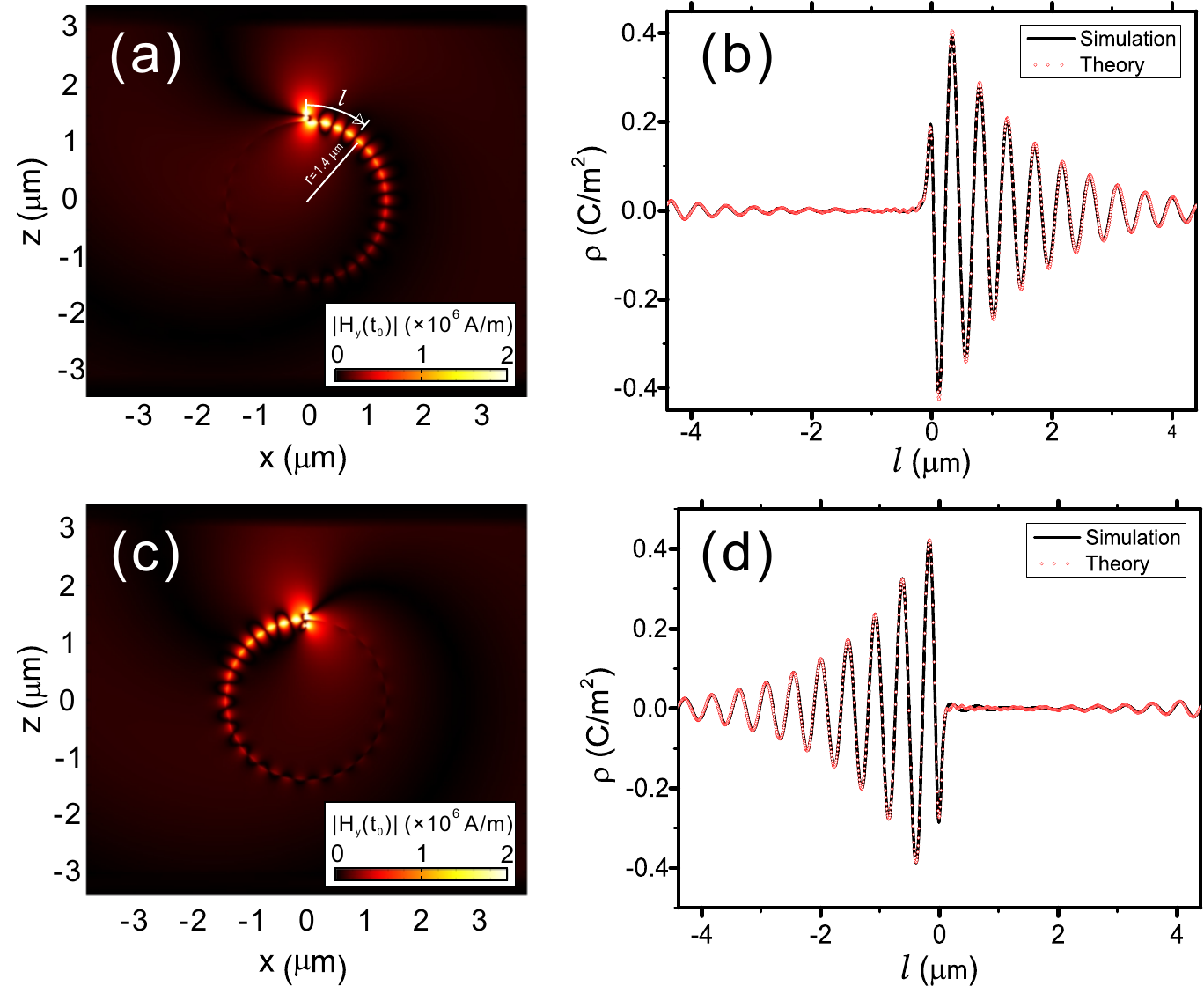}
\caption{\label{fig4} Directional excitation of GPs on curved free-standing graphene. The distribution of magnetic field $|\text{Re}\{H_z\}|$ for GPs excited by the configuration of case \uppercase\expandafter{\romannumeral1}(a) and two separated orthogonal polarized dipoles (c). Simulated and analytically calculated spatial dependent charge density for the case \uppercase\expandafter{\romannumeral1}(b) and case \uppercase\expandafter{\romannumeral2}(d). The phase in (c) is set as $\pi$/4 in order to show the two individual sources.}
\end{figure}

Except for the scenario of directional excitation of GPs in a flat graphene plane, the directional generation of GPs on a curved free-standing graphene sheet is also investigated. The curvature breaks the mirror symmetry relative to graphene plane, then the induced radiative loss will affect the excitation and propagation efficiency of GPs. The critical curvature radius which permits confined wave exist can be calculated by $r_c\approx\frac{k_0}{\text{Im}\{k_z\}(k_\text{spp}-k_0)}\approx0.048\lambda_\text{spp}$,  thus a circular radii as $r=1.4~\mu m \approx 3 \lambda_\text{spp}$ is chosen, which is a typical value in flexible transformation plasmonics\cite{LZX2013}. A circularly polarized dipole with momentum as the same as in flat graphene is placed above the graphene circle at a distance of 100~nm. The configuration and simulated $H_y$ field amplitude are shown in Fig. \ref{fig4}(a). Remarkably, the mode propagates mainly along clockwise direction. which is in coincide with the result in flat graphene (case \uppercase\expandafter{\romannumeral1}). We can describe the induced charge density by
\begin{equation}\label{eqrho}
    \rho(l)\approx\rho^f(x)+\rho^f(x\mp2\pi r), ~l\in(-\pi r, \pi r]
\end{equation}
the upper (lower) sign in Eq. \ref{eqrho} applies to $l>0$ ($l<0$), where $l=r\theta$ is arc length away from the dipole. The spatial charge density ratio is shown in Fig. \ref{fig4}(b). One can know that the directional excitation behavior of GPs in arc surface can be understood well by flat graphene with the same parameter. When the circularly polarized dipole is decomposed into two dipoles located both above and below the graphene, the simulated $H_y$ field amplitude shown in \ref{fig4}(c) and charge density distribution shown in \ref{fig4}(d) can be understood well from flat graphene in configuration of case \uppercase\expandafter{\romannumeral2}. These results show that directional propagation of GPs can be extended into arc surfaces directly.
\begin{figure}[htb]
\includegraphics[width=8.5cm]{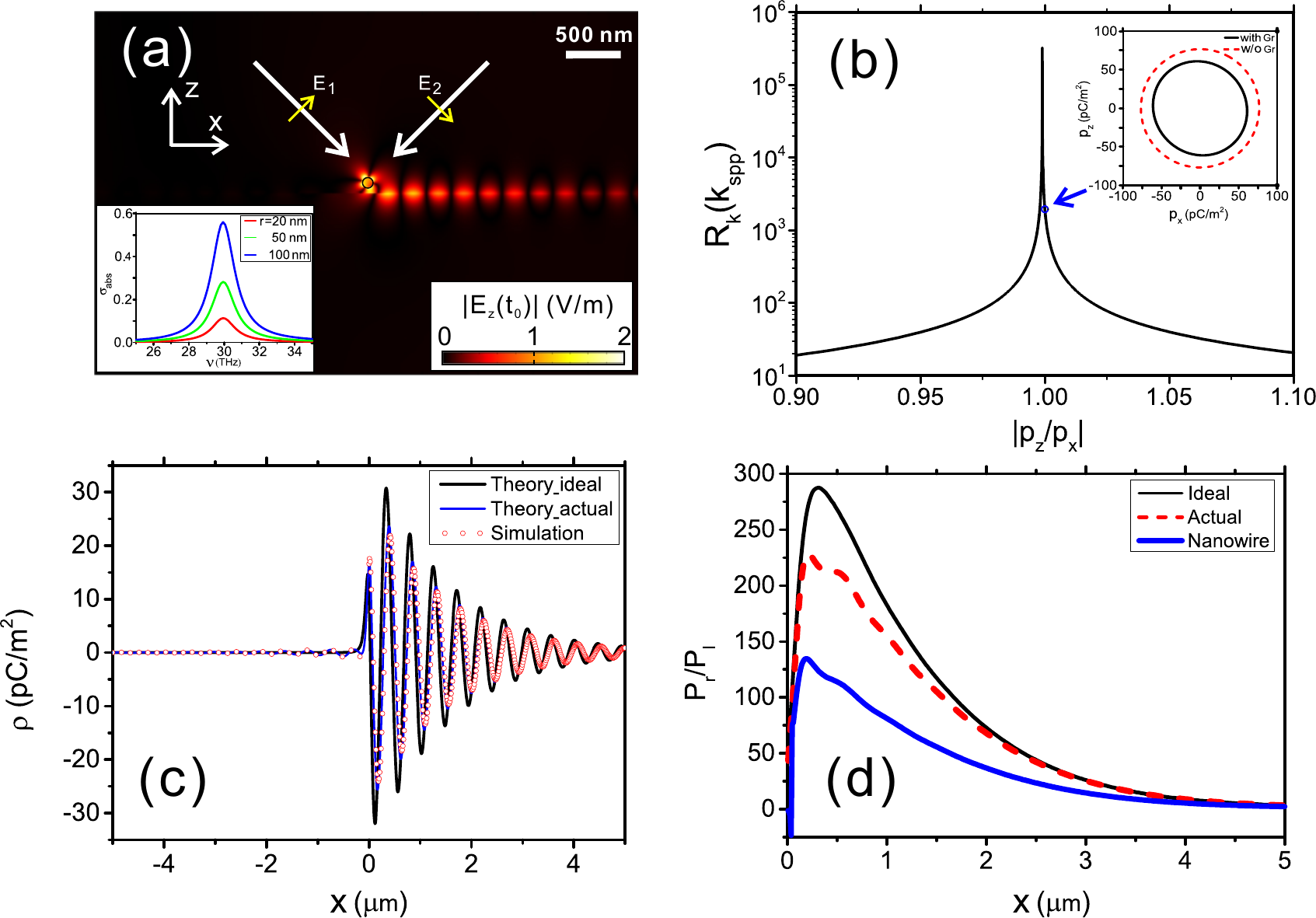}
\caption{\label{fig5}Directional excitation of GPs using $\text{In}_{0.53}\text{Ga}_{0.47}\text{As}$ nanowires illuminated by two orthogonally polarized plane waves. (a) Schematics representation and electric field distributions $|E_z|$ for excited GPs, the insert figure is normalized absorption cross section of the $\text{In}_{0.53}\text{Ga}_{0.47}\text{As}$ nanowire with radii of 20~nm, 50~nm and 100~nm. (b) The dependence of the angular spectrum ratios $R_k(k_\text{spp})$ on extinction parameter $|p_z/p_x|$. The blue circle indicates the parameter of the considered nanowire. The insert figure is the polarized circle of the nanowire with (solid line) and without (broken line) graphene sheet. (c) The simulated and theoretically calculated spatial dependent charge density in graphene. The thick line indicates the charge distribution induced by ideal circular polarized dipole with dipole momentum as $-i\omega p_x=76.72~\text{pA}\cdot \text{m}$, the thin line indicate the case of dipole momentum as $-i\omega p_x=61.08~\text{pA}\cdot \text{m}$ and $\frac{p_z}{p_x}=-0.0624+0.9953i$, respectively. The dot marked line indicates the simulated result of semiconductor nanowire. (d) The energy flux ratio of ideal dipole ($\frac{p_z}{p_x}=i$, solid line), actual dipole($\frac{p_z}{p_x}=-0.0624+0.9953i$, broken line) and simulated result of semiconductor nanowire (thick line).}
\end{figure}

Next, we turn to discuss how to realize our proposal in real experiments. The dipole employed in the paper can be mimicked by a semiconductor nanowire illuminated by two orthogonally polarized plane waves. Due to the wave interference, the background standing wave satisfies $E_x\propto\sqrt{2}\cos(k_xx+\pi/4)$, $E_z\propto\sqrt{2}i\sin(k_xx+\pi/4)$ and $p_z/p_x=E_z/E_x=i\tan(k_xx+\pi/4)$, thus the amplitude condition of ideal circular dipole requires $k_xD\ll\pi/4$, where $D$ is the dimension of nanowire. In the configuration of directional excitation of metallic plasmons, $k_xD\sim1/\sqrt{2}$, the induced dipole  moment is too nonuniform to be treated as ideal circular dipole source. That is to say, this method is not suitable for directional excitation of metallic plasmons. However, this is not a limitation any more in excitation of GPs due to the deep sub wavelength of the nanowire size in infrared spectrum, ie, $k_xD\ll1/\sqrt{2}$. In our simulation, an $\text{In}_{0.53}\text{Ga}_{0.47}\text{As}$ nanowire with diameter $D$ of 100~nm (0.01$\lambda$) is used to mimic the rotational polarized dipole. Drude model was adopted to model the dielectric constant of $\text{In}_{0.53}\text{Ga}_{0.47}\text{As}$. In this model, the dielectric function is given by $\epsilon(\omega)=\epsilon_\infty-\frac{\omega_p^2}{\omega(\omega+i\gamma)}$ , where $\omega_p=\sqrt{nq^2/m^{*}_e\epsilon_0}$ is the plasma frequency, $\epsilon_\infty$ is the high frequency dielectric constant, and $\gamma=q/\mu_em_e^*$ is damping rate. Extracted from the reference in\cite{HAH2007}, the parameters of $\text{In}_{0.53}\text{Ga}_{0.47}\text{As}$ are $\epsilon_\infty=12.15$, $\tau=\gamma^{-1}=0.1$~ps, and $m^{*}=0.523m_e$. Moreover, $n=6.3\times10^{18}/\text{cm}^3$ is used to realize the resonance of the nanowire at 30~THz. The absorption cross length normalized to geometry cross length for different diameters of the nanowire are showed in the insert figure of Fig. \ref{fig5}(a), the absorption peak lays at 30~THz and is independent on the diameter of the nanowire because the electrostatic approximation is satisfied. Two time harmonic orthogonal incident plane waves with amplitude of 1~V/m and phase difference of $\pi/2$ are taken to illuminate the wire, the schematic and simulated electric field distribution are depicted in Fig. \ref{fig5}(a), where the incident field has been subtracted from the total field. The expression of the incident fields adopted in the simulation is expressed as
\begin{eqnarray*}
  \mathbf{E}_1&=&\frac{1}{\sqrt{2}}(\hat{\bm{x}}+\hat{\bm{z}})e^{ik_xx-ik_zz-i\omega t} \\
  \mathbf{E}_2&=&\frac{1}{\sqrt{2}}(\hat{\bm{x}}-\hat{\bm{z}})e^{-ik_xx-ik_zz-i\omega t-i\pi/2}.
\end{eqnarray*}
It would be expected that the nanowire serves as a circularly polarized dipole with $p_z/p_x=i$. From the figure, one can see that induced near field along $+x$ with a much larger amplitude than the one along $-x$, which is very similar to the case of a circularly polarized dipole with $p_z/p_x=i$.  The induced dipole of the nanowire in the diameter of 100~nm is  $-i\omega p_x=76.72~\text{pA}\cdot \text{m}$ and $\frac{p_z}{p_x}=i$, which means that the semiconductor nanowire can serve as an ideal circularly polarized dipole as expected. When a graphene sheet is introduced close to this nanowire, the electric field reflected from the graphene will act on the nanowire as well, and this changes the parameters of the induced dipole to $-i\omega p_x=61.08~\text{pA}\cdot \text{m}$ and $\frac{p_z}{p_x}=-0.0624+0.9953i$, respectively. The vibration ellipses of the induced dipole with and without graphene are shown in the insert figure of Fig. \ref{fig5}(b). From the angular momentum ratio shown in Fig. \ref{fig5}(b), the angular momentum ratio can over 1000 for the nanowire. The induced charge density distribution is shown in Fig. \ref{fig5}(c), one can find that the charges oscillate only in the $+x$ direction. The simulated charge distribution is in good agreement for analytical calculation when the dipole moment is set as the actual value of $-i\omega p_x=61.08~\text{pA}\cdot \text{m}$ and $\frac{p_z}{p_x}=-0.0624+0.9953i$, respectively. The charge distribution of unperturbed ideal circular polarized dipole is shown in thick line, one can see that the oscillation amplitude is a bit larger than actual situation. The energy flux ratios are shown in the Fig. \ref{fig5}(d), the asymmetrical energy flux is very apparent, the unperturbed ideal result is given for comparison as well. The energy flux ratio with extinction parameter of $\frac{p_z}{p_x}=-0.0624+0.9953i$ is similar to the ideal case except for small extra oscillation and less magnitude due to the existence of real part of the extinction parameter. The simulated result of nanowire is similar to the analytical result with actual extinction parameter, one can see that the energy flux ratio mimicked by nanowire exceed 100 when the propagation length is less than 2$\lambda_\text{spp}$, and the extinction ratio exceed 10 in the whole calculation window.



\section{\label{sec:level4}CONCLUSION}
We demonstrated here that near field interference of a circularly polarized dipole and two mirror image symmetric dipoles with orthogonally polarizations can directional generate propagating GPs. The viewpoint of angular momentum conservation is very efficient to determine the preferred direction of exited GPs. When the dipoles are laid in different sides of graphene, the spatial charge density rather than the magnetic field should be adopted to analysis the excited GPs due to the extra minus sign from the discontinuous of magnetic field. In this condition, the magnetic field of dipole and induced charge distribution have opposite preferred directions and the properties of excited GPs should be described by the behavior of induced charge. Moreover, the direction generation of GPs can be extended into arc surface directly. Furthermore, a semiconductor nanowire can be regarded as a localized source to mimic the polarized dipoles, which can be realized in real experiments.
\begin{acknowledgments}
W Cai, X Zhang, and J Xu acknowledge support from the National Basic Research Program of China (2013CB328702), Program for Changjiang Scholars and Innovative Research Team in University (IRT0149), the National Natural Science Foundation of China (11374006) and the 111 Project (B07013). Y. Luo acknowledge support from the National Natural Science Foundation of China (61574122).
\end{acknowledgments}


\begin{thebibliography}{27}
\bibitem[]{sup}{The even parity in angular spectrum originate from $\rho^{p_z}(-k_x)=\rho^{p_z}(k_x)$, and the even parity in real space is result from
$\rho^{p_z}(-x)=\int_{-\infty}^\infty\rho^{p_z}(k_x)e^{-ik_xx}dk_x=\int_{-\infty}^\infty \rho^{p_z}(-k_x)e^{ik_xx}dk_x=\rho^{p_z}(x)$. }
\end{thebibliography}

 %
\end{document}